\documentclass[pre,10pt,twocolumn,preprint]{revtex4}
\usepackage[dvips]{graphicx}
\usepackage{latexsym}
\usepackage{natbib}
\usepackage[centerlast]{caption}
\usepackage{subfig}
\usepackage{color}
\definecolor{red}{rgb}{1,0,0}
\usepackage{amsmath}
\makeatletter

\newcommand{\Rmnum}[1]{\expandafter\@slowromancap\romannumeral#1@}
\makeatother

\begin{document}

\title{Robustness of a Network of Networks}

\author{Jianxi Gao,$^{1,2}$ Sergey V. Buldyrev,$^3$ Shlomo Havlin,$^4$ and
H. Eugene Stanley$^1$}

\affiliation{$^1$Center for Polymer Studies and Department of
Physics, Boston University, Boston, MA 02215 USA\\
$^2$Department of Automation, Shanghai Jiao Tong University, 800
Dongchuan Road, Shanghai 200240, PR China\\
$^3$Department of Physics,~Yeshiva University, New York, NY 10033 USA\\
$^4$Minerva Center and Department of Physics, Bar-Ilan University,
52900 Ramat-Gan, Israel}

\date{25 October 2010 --- gbhs25oct.tex}

\begin{abstract}

Almost all network research has been focused on the properties of a
single network that does not interact and depends on other networks.
In reality, many real-world networks interact with other networks.
Here we develop an analytical framework for studying interacting
networks and present an exact percolation law for a network of $n$
interdependent networks. In particular, we find that for $n$
Erd\H{o}s-R\'{e}nyi networks each of average degree $k$, the giant
component, $P_{\infty}$, is given by
$P_{\infty}=p[1-\exp(-kP_{\infty})]^n$ where $1-p$ is the initial
fraction of removed nodes. Our general result coincides for $n=1$
with the known Erd\H{o}s-R\'{e}nyi second-order phase transition for
a single network. For any $n \geq 2$ cascading failures occur and
the transition becomes a first-order percolation transition. The new
law for $P_{\infty}$ shows that percolation theory that is
extensively studied in physics and mathematics is a limiting case
($n=1$) of a more general general and different percolation law for
interdependent networks.

\end{abstract}

\maketitle In recent years dramatic advances in the field of complex
networks have occurred
~\cite{Strogatz1998,bara2000,Callaway2000,Albert2002,Cohen2000,Newman2003,
Dorogovtsev2003,song2005,Pastor2006,Caldarelli2007,Barrat2008,Shlomo2010,Neman2010}.
The internet, airline routes, and electric power grids are all
examples of networks whose function relies crucially on the
connectivity between the network components. An important property
of such systems is their robustness to node failures. Almost all
research has been concentrated on the case of a single or isolated
network which does not interact with other networks. Recently, based
on the motivation that modern infrastructures are becoming
significantly more dependent on each other, a system of two coupled
interdependent networks has been studied~\cite{Sergey2010}. A
fundamental property of interdependent networks is that when nodes
in one network fail, they may lead to the failure of dependent nodes
in other networks which may cause further damage in the first
network and so on, leading to a global cascade of failures. Buldyrev
et al.~\cite{Sergey2010} developed a framework for analyzing
robustness of two interacting networks subject to such cascading
failures. They found that interdependent networks become
significantly more vulnerable compared to their noninteracting
counterparts.

For many important examples, more than two networks depend on each
other. For example, diverse infrastructures are coupled together,
such as water and food supply, communications, fuel, financial
transactions, and power
stations~\cite{Peerenboom2001,Rinaldi2001,Rosato2008,Alessandro2010}.
For further examples see Section \Rmnum{1} in the Supplementary
Information (SI). Understanding the robustness due to such
interdependencies is one of the major challenges for designing
resilient infrastructures.

We introduce here a model system, comprising a network of $n$
coupled networks, where each network consists of $N$ nodes (See Fig.
1). The $N$ nodes in each network are connected to nodes in
neighboring networks by bidirectional dependency links, thereby
establishing a one-to-one correspondence as illustrated in Fig. 2 in
SI. We develop a mathematical framework to study the robustness of
this ``network of networks'' (NON). We find an exact analytical law
for percolation of a NON system composed of $n$ coupled randomly
connected networks. Our result generalizes the known
Erd\H{o}s-R\'{e}nyi (ER)~\cite{ER1959,ER1960,Bollob1985} result for
the giant component of a single network and the $n=2$ result found
recently~\cite{Sergey2010}, and shows that while for $n=1$ the
percolation transition is a second order transition, for $n>1$
cascading failures occur and the transition becomes a first order
transition. Our results for $n$ interdependent networks suggest that
the classical percolation theory extensively studied in physics and
mathematics is a limiting case of a general theory of percolation in
NONs, or networks with multiple types of connectivity links. This
general theory has many novel features which are not present in
classical percolation.

Additionally, we find:

(i) the robustness of NON {\it significantly decreases\/} with $n$, and

(ii) for a network of $n$ ER networks all with the same average
degree $k$, there exists a minimum degree $k_{\min}(n)$ increasing
with $n$, below which $p_c =1$, i.e., for $k<k_{\min}$ the NON will
collapse once any finite number of nodes fail. We find an analytical
expression for $k_{min}(n)$, which generalizes the known result
$k_{\min}(1)=1$ for ER below which the network collapses. We also
discuss the critical effect of loops in the NON structure.

Real-world interacting networks (See SI for more details)
are characterized by complex
correlations and a variety of organizational principles governing their
internal structures and interdependencies. Once these correlations are
quantified from the statistical analysis of actual data bases and the
organizational principles are specified from the engineering literature,
real world networks can be studied by computer simulations. These
simulations will have many parameters and therefore their outcome will
also require complex interpretation.  It is therefore very important to
develop simple analytically tractable models for the robustness of
interdependent networks against which such simulations can be
tested. Well-known examples of simplified models that both demonstrate a
fundamental phenomenon and significantly advance our knowledge are the
Ising model in statistical mechanics and the Erd\H{o}s-R\'{e}nyi model
in graph theory. This paper presents a simple model that can serve as a
benchmark for further studies of NONs.

We assume that a network $m$ $(m=1,2,...,n)$ in the NON is a
randomly connected network with a degree distribution $P_m(k)$. We
call a pair of networks A and B a fully interdependent pair if it
satisfies the following condition: each node $A_i$ of network A is
connected to one and only one node $B_i$ in network B by a
bidirectional dependency link such that if node $A_i$ fails, $B_i$
also fails and vice versa. Since the number of nodes in each network
is the same, these bidirectional links establish a one-to-one
correspondence between the nodes in the networks belonging to an
interdependent pair. Each node of the NON represents a network and
each edge represents a fully interdependent pair of networks. First,
we will discuss the case when the NON is a loopless tree of $n$
networks (Fig. 1). The dependency edges in such a NON establish a
unique one-to-one correspondence between the nodes of any two
networks not necessarily belonging to the same fully interdependent
pair. This one-to-one correspondence established by the
interdependency links between the nodes of different networks in the
loopless NON uniquely maps any set of nodes in one of the networks
to a set of nodes (which we will call an image of the original set)
in any other network of the NON (See SI for more details). In
principle, the assumption of full interdependence allows one to
collapse all the networks of the NON onto a single network with
multiple types of links.

We assume that in order to remain functional a node must belong to a
sufficiently large mutually connected cluster~\cite{Sergey2010} (See
detailed definition in SI). We will show that a large mutually
connected cluster which includes a finite fraction of the nodes in
each network exists only in networks of sufficiently high mean
degree. We call this large mutually connected cluster a mutual giant
component, and we postulate that only nodes in the mutual giant
component remain functional.

We assume that due to an attack or random failure only a fraction of
nodes $p$ in one particular network which we will call the root of
the NON. We can now observe a cascade of failures caused by the
failure of the dependent nodes in the networks connected directly to
the root by the edges of the NON. The damage will further spread to
more distant networks. Moreover, fragmentation of each network
caused by the removal of certain nodes will cause malfunction of
other nodes which will now belong to small isolated clusters. This
malfunction will cause dependent nodes in neighboring networks to
malfunction as well. Depending on the time scales of these
processes, the damage can spread across the NON back and forth,
which we can visualize as cascades of failures, as shown in Fig. 3
of the SI section. At the end of this process only the mutual giant
component of the NON, if it exists, will remain functional.

We now introduce generating
functions~\cite{Sergey2010,Newman2001,Newman2002PRE,Shao2008,Shao2009}
of each network, $G_{m0}(z)=\sum P_m(k)z^k$, and the generating
function of the associated branching process, $G_{m1}(z)=G'_{m0}(z)/
G'_{m0}(1)$. It is known~\cite{Newman2001,Newman2002PRE} that the
generating functions of a randomly connected network once a fraction
$1-p$ of nodes has been randomly removed are the same as the
generating functions of the original network with the new argument
$1-p(1-z)$. Furthermore it is known~\cite{Shao2008,Shao2009} that
the fraction of nodes in the giant component of a single randomly
connected network is $\mu_{\infty,1}=pg_m(p)$, where
$g_m(p)=1-G_{m0}(1-p(1-f_m))$ and $f_m$ satisfies a transcendental
equation $f_m(p)=G_{m1}(1-p(1-f_m(p)))$.

We next prove that the fraction of nodes, $\mu_{\infty,n}$, in the
mutual giant component of a NON composed of $n$ networks is the
product:
\begin{equation}
\mu_{\infty,n}=p\prod_{m=1}^n g_m(x_m), \label{e:mu}
\end{equation}
where each $x_m$ satisfies the equation
\begin{equation}
x_m=\mu_{\infty,n}/g_m(x_m). \label{e:x}
\end{equation}
The system of $n+1$ equations (\ref{e:mu}) and (\ref{e:x}) defines
$n+1$ unknowns: $ \mu_{\infty,n}, x_1, x_2, ..., x_n$ as functions
of $p$ and the degree distributions $P_m(k)$.

We derive Eqs. (1) and (2) by mathematical induction. (An
alternative proof is given in the SI). Indeed, for $n=1$, Eqs. (1)
and (2) follow directly from the definition of $\mu_{\infty,1}$.
Assuming that the formulas are valid for a NON of $n-1$ networks we
will prove that they are valid also for a NON of $n$ networks. A
loopless NON of $n$ networks can be always represented as one of its
networks connected by a single edge to the other $n-1$ networks in
the NON. All the nodes in the $n$-th network, which do not belong to
the image of the mutual giant component $\mu_{\infty,n-1}$ of the
NON of $n-1$ networks will stop to function. The fraction of the
nodes in the image of this mutual giant component onto the $n$-th
network satisfies the equation $x_{1,n}=\mu_{\infty,n-1}(p)$. The
fraction of nodes belonging to the giant component of this
dependency image is $\mu_{1,n} = x_{1,n} g_n(x_{1,n})$. Only the
nodes in the NON of $n-1$ networks which belong to the dependency
image of the giant component of the $n$-th network will remain
functional. Due to the randomness of the connectivity links in
different networks, this dependency image can be represented as a
random selection of the fraction $g_n(x_{1,n})$ out of the
originally survived nodes, or as random selection of $p_1=p
g_n(x_{1,n})$ fraction of nodes in one of the networks comprising
the NON of $n-1$ networks. The fraction of nodes in the new mutual
giant component of the NON of $n-1$ networks corresponding to this
new random selection will be $\mu_{\infty,n-1}(p_1)$. The image of
this mutual giant component in the $n$-th network is equivalent to a
random selection of $x_{2,n}=\mu_{\infty,n-1}(p_1)/g_n(x_{1,n})$
fraction of nodes out of the entire $n$-th network. As we continue
this process, the sequence of giant components $\mu_{j,n}$ in the
$n$-th network, randomly selected sets $x_n$ in the $n$-th network
and randomly selected sets $p_j$ in the NON of $n-1$ networks will
satisfy the recursion relations $x_{j+1,n}
=\mu_{\infty,n-1}(p_j)/g_n(x_{j,n})$,
 $\mu_{j+1,n}=x_{j+1,n}g_n(x_{j+1,n})$,
$p_{j+1}=p g_n(x_{j+1,n})$.

In the limit $j\to \infty$, this process will converge, i.e. all the
parameters in the two successive steps will coincide: $x_{j+1,n}\to
x_{j,n} \equiv x_n$, $p_j \to p g_n(x_n)$ and $\mu_{\infty,n-1}(p_j)
\to \mu_{\infty,n}$. Then $x_n=\mu_{\infty,n}/g_n(x_n)$ which is
identical to the last equation in Eqs. (2) and
$\mu_{\infty,n-1}(p_j) \to p
g_n(x_n)\prod_{m=1}^{n-1}g_m(x_m)=p\prod_{m=1}^n g_m(x_m) \equiv
\mu_{\infty,n}$ which is identical to Eq. (1). By the assumption of
induction $x_m=\mu_{\infty,n-1}(p_j)/g_m(x_m)=
\mu_{\infty,n}/g_m(x_m)$ which completes the set of Eqs. (2).
Finally $\mu_{j+1,n}\to x_n g_n(x_n)=\mu_{\infty,n}/g_n(x_n)$ and
thus the fraction of nodes in the giant $n$-th network coincides
with the mutual giant component in the NON of $n-1$ networks. The SI
presents an alternative analytical derivation of Eqs. (1) and (2),
which represent a certain type of cascading failures. The SI also
presents simulation results which agree well with the theory (Figs.
5 and 6 in SI).

For the case of a NON with loops, the closed path of fully
interdependent pairs starting form a network A and ending at the
same network A will establish a dependence of nodes $A_i$ on node
$A_{j_i}$, where $j_i$ is a transposition of $i$. Then the failure
of single node $i$ will cause an entire cycle in the transposition
to fail. The average size of a cycle in the transposition of $N$
elements grows as $N/\ln N$, so the initial failure of $\ln N$
nodes will cause almost all the nodes of the NON to fail without
taking into account any connectivity links which will cause
additional damage. So the NON with loops is unstable against removal
of an infinitely small fraction of nodes unless the transposition
$j_i$ is not random. In case when the transposition $j_i$ is
trivial, $j_i=i$, we have the same one-to-one correspondence between
the nodes as in the loopless NON and then Eq. (\ref{e:mu}) and
(\ref{e:x}) are valid. This is since in our proof we did not use any
other property of a NON except the unique one-to-one correspondence
of the nodes in different networks.

The case of NON of $n$ Erd\H{o}s-R\'{e}nyi
(ER)~\cite{ER1959,ER1960,Bollob1985} networks with average degrees
$k_1, k_2,... k_i,...,k_n$ can be solved explicitly. In this case,
we have $G_{1,i}(x)=G_{0,i}(x)=\exp[k_i(x-1)]$ \cite{Newman2002PRE}.
Accordingly $g_i(x_i) =1-\exp[k_i x_i(f_i-1)]$, where $f_i=\exp[k_i
x_i(f_i-1)]$ and thus $g_i(x_i)=1-f_i$. Using Eq. (\ref{e:x}) for
$x_i$ we get
\begin{equation}\label{eq3}
f_{i} = \exp[-pk_i \prod_{j=1}^{n} (1-f_j)], i = 1,2,...,n.
\end{equation}
These equations can be solved analytically, as shown in detail in
the SI section. They have only a trivial solution ($f_i=1$) if
$p<p_c$, where $p_c$ is the mutual percolation threshold. When the
$n$ networks have the same average degree $k$, $k_i=k$
($i=1,2,...,n$), we obtain from Eq. (3) that $f_c \equiv f_i(p_c)$
satisfies
\begin{equation}\label{eq7}
f_c = e^{\frac{f_c-1}{nf_c}}.
\end{equation}
where the solution can be expressed in term of the Lambert function
$W(x)$~\cite{Lambert1758,Corless1996}, $f_c =
-[nW(-\frac{1}{n}e^{-\frac{1}{n}})]^{-1}$.

Once $f_c$ is known, we obtain $p_c$ and $\mu_{\infty,n} \equiv
P_{\infty}$ by substituting $k_i=k$ into Eq. (S10) of the SI section
\begin{equation}\label{equ8}
\begin{array}{lcl}
p_c = [nkf_c(1-f_c)^{(n-1)}]^{-1},& \mbox{} & \\ P_{\infty} =
\frac{1-f_c}{nkf_c}.& \mbox{} & \\
\end{array}.
\end{equation}
For $n=1$ we obtain the known result $p_c=1/k$ of
Erd\H{o}s-R\'{e}nyi \cite{ER1959,ER1960,Bollob1985}. Substituting
$n=2$ in Eqs. (\ref{eq7}) and (\ref{equ8}) one obtains the exact
results of \cite{Sergey2010}.

To analyze $p_c$ as a function of $n$, we find $f_c$ from Eq.
(\ref{eq7}) and substitute it into Eq. (\ref{equ8}), and we obtain
$p_c$ as a function of $n$ for different $k$ values, as shown in Fig.~2(a).
It is seen that the NON becomes more vulnerable with increasing $n$
or decreasing $k$ ($p_c$ increases when $n$ increases or $k$
decreases). Furthermore, for a fixed $n$, when $k$ is smaller than a
critical number $k_{min}(n)$, $p_c \geq 1$ meaning that for
$k<k_{min}(n)$, the NON will collapse even if a single node fails.
Fig.~2 (b) shows the minimum average degree $k_{\min}$ as a function
of the number of networks $n$. From Eq. (\ref{equ8}) we get the
minimum of $k$ as a function of $n$
\begin{equation}\label{eq10}
k_{\min}(n) = [nf_c(1-f_c)^{(n-1)}]^{-1},
\end{equation}
Note that Eq. (\ref{eq10}) together with Eq. (\ref{eq7}) yield the
value of $k_{\min}(1)=1$, reproducing the known ER result, that
$\langle k \rangle =1$ is the minimum average degree needed to have
a giant component. For $n=2$, Eq. (\ref{eq10}) yields the result
obtained in \cite{Sergey2010}, i.e., $k_{\min}=2.4554$.

From Eqs. (1)-(3) we obtain the percolation law for the order
parameter, the size of the mutual giant component for all $p$ values
and for all $k$ and $n$,
\begin{equation}\label{eq11}
\mu_{\infty,n} \equiv P_{\infty} = p[1-\exp(-kP_{\infty})]^n.
\end{equation}
The solutions of equation (\ref{eq11}) are shown in Fig. 3 for
several values of $n$. Results are in excellent agreement with
simulations. The special case $n=1$ is the known ER percolation law
for a single network \cite{ER1959,ER1960,Bollob1985}. Note that
Eqs.~(4)--(7) are based on the assumption that all $n$ networks have
the same average degree $k$.

In summary, we have developed a framework, Eqs. (1) and (2), for
studying percolation of NON from which we derived an exact
analytical law, Eq.~(\ref{eq11}), for percolation in the case of a
network of $n$ coupled ER networks. Equation~(\ref{eq11}) represents
a bound for the case of partially dependent networks
\cite{parshani2010}, which will be more robust. In particular for
any $n \geq 2$, cascades of failures naturally appear and the phase
transition becomes first order transition compared to a second order
transition in the classical percolation of a single network. These
findings show that the percolation theory of a single network is a
limiting case of a more general case of percolation of
interdependent networks. Due to cascading failures which increase
with $n$, vulnerability significantly increases with $n$. We also
find that for any loopless network of networks the critical
percolation threshold and the mutual giant component depend only on
the number of networks and not on the topology (see Fig.~1(a)). When
the NON includes loops, and dependency links are random, $p_c=1$ and
no mutual giant component exists.

\begin{figure}[h!]
\centering \includegraphics[width=0.44\textwidth]{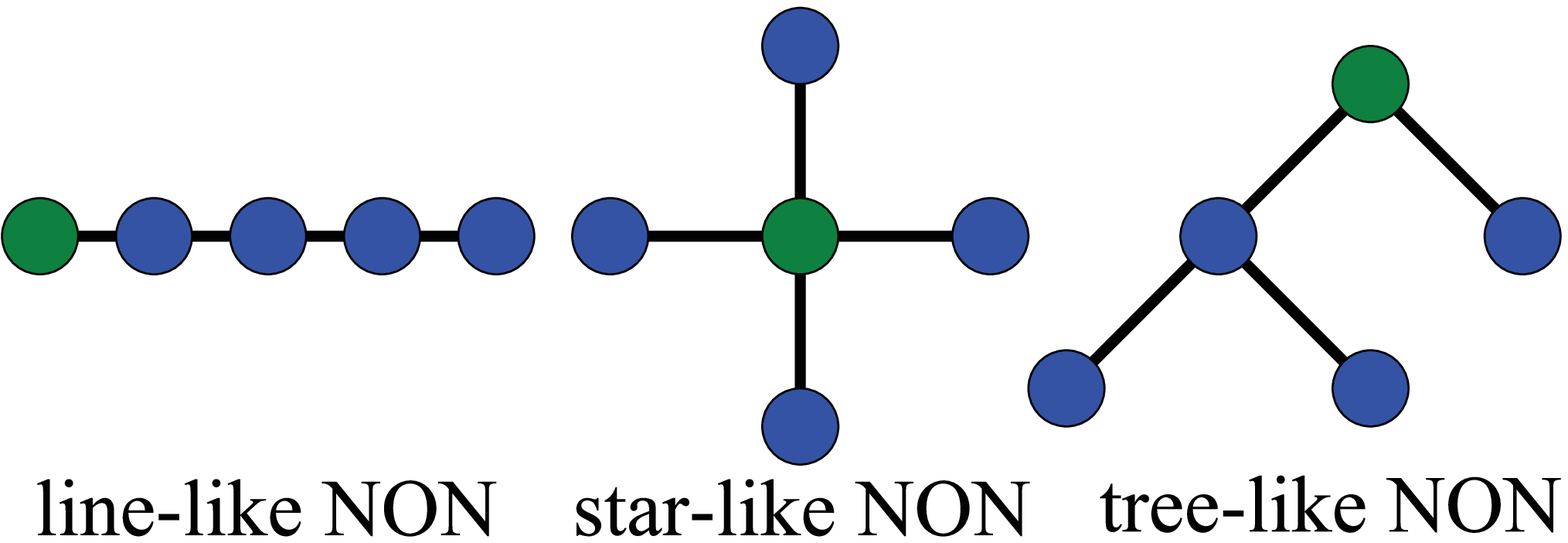} \\[20pt]
\caption{(color online)  Three types of loopless networks of
networks composed of five coupled networks. All have same
percolation threshold and same giant component. The darker green
node is the origin network on which failures occur. }
\end{figure}\label{fig1}

\begin{figure}[h!]
\centering \includegraphics[width=0.44\textwidth]{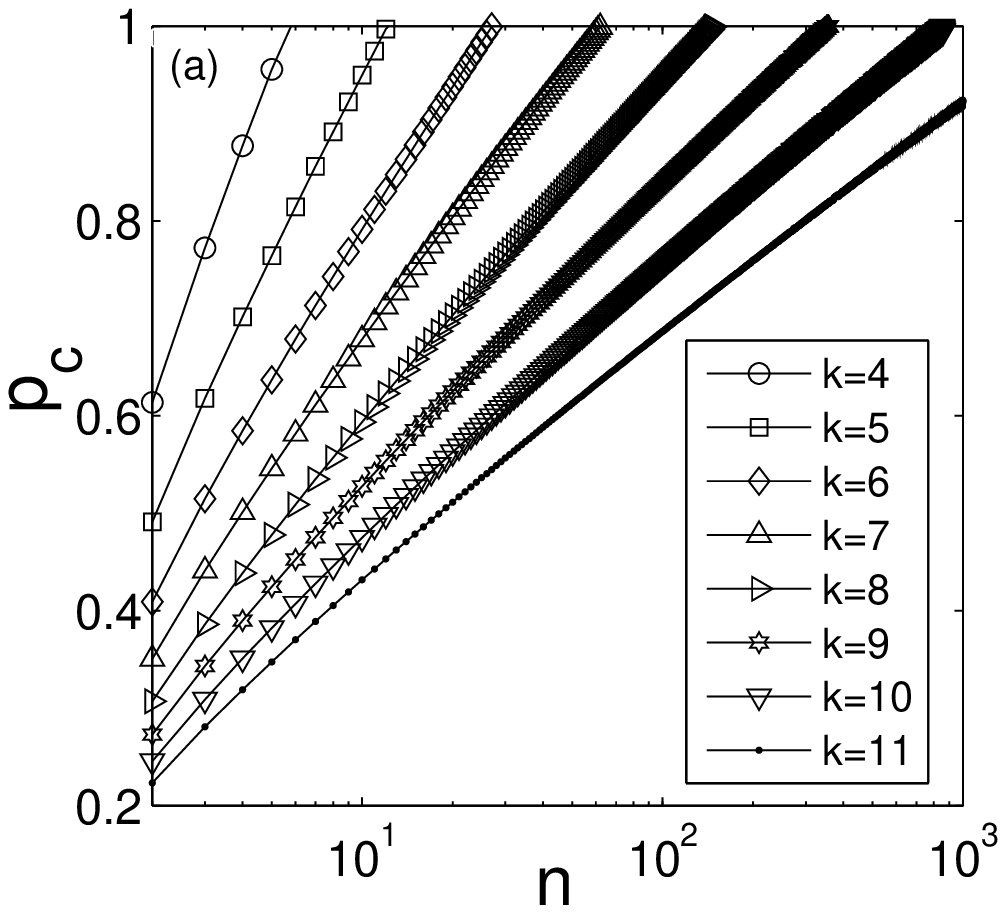}
\hspace{.0in} \includegraphics[width=0.44\textwidth]{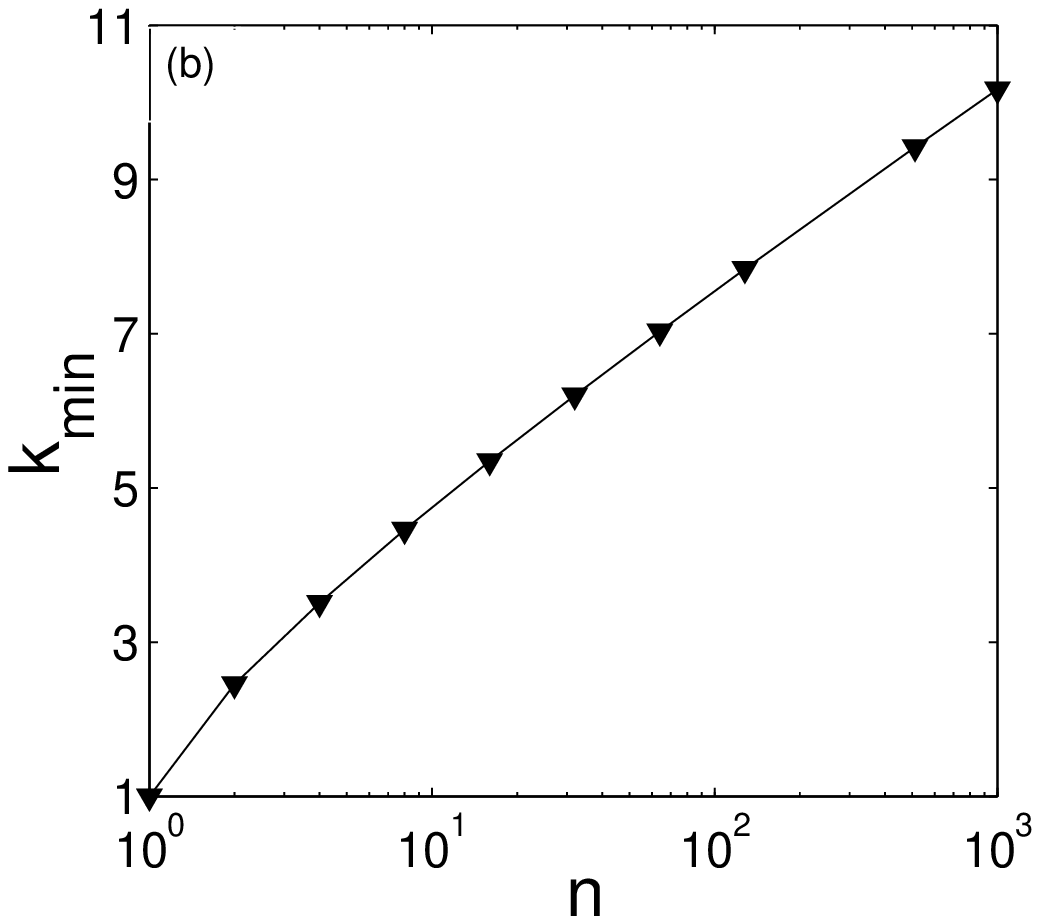}
\caption{(a) The critical fraction $p_c$ for different $k$ and $n$
and (b) minimum average degree $k_{\min}$ as a function of the
number of networks $n$. The results of (a) and (b) are obtained
using Eqs. (\ref{equ8}) and (\ref{eq10}) respectively and are in
good agreement with simulations. In simulations $p_c$ was calculated
from the number of cascading failures which diverge at $p_c$
\cite{parshani2010} (see also Fig. 7 in SI).}\label{fig5}
\end{figure}\label{fig2}

\begin{figure}[h!]
\centering \includegraphics[width=0.44\textwidth]{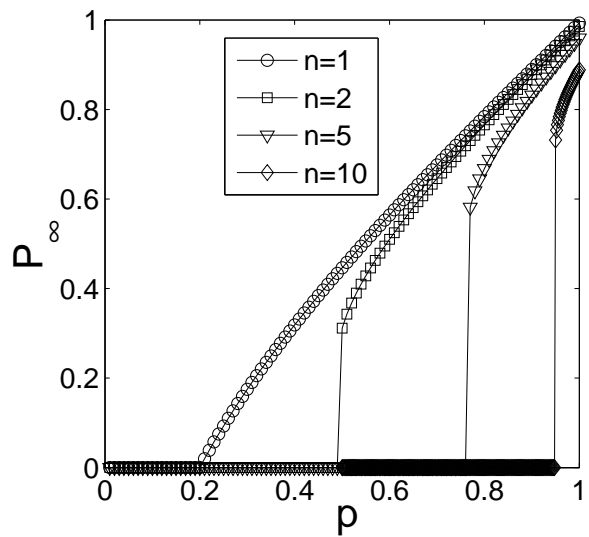}
\caption{ For loopless NON, $P_{\infty}$ as a function of $p$ for
$k=5$ and several values of $n$. The results obtained using Eq.
(\ref{eq11}) agree well with simulations.}\label{fig3}
\end{figure}

\end{document}